\documentstyle[12pt]{article}
\date{}
\newcommand{\half}{{1\over2}}
\def\r{{\mbox{{\protect\scriptsize r}}}}
\def\eff{{\mbox{{\protect\scriptsize eff}}}}
\def\Tro{T_{\r}}
\def\Teff{T_{\eff}}
\newcommand{\str}{\rule{0ex}{2.7ex}}
\newcommand{\mc}{\multicolumn}
\title{%
  \vskip-4em\hfill {\small \begin{tabular}[t]{l}
                       \rule{0ex}{1ex} CERN-TH.6893/93 \\[.0ex]
                       \rule{0ex}{1ex} FSU-SCRI-93-67  \\[.0ex]
                       \rule{0ex}{1ex} cond-mat/9305029\\[.0ex]
                       \rule{0ex}{1ex} May 1993
  \end{tabular} \break\vskip3em  }
\Large\bf
The Solid-on-Solid Surface Width Around the Roughening Transition}
\author{\vbox{\vspace{0mm}}\\[-4mm]%
 Hans Gerd Evertz$^1$, Martin Hasenbusch$^2$,\\
 Mihai Marcu$^3$ and  Klaus Pinn$^4$\\[6mm]
\small
  $^1\,$Supercomputer Computations Research Institute \\[-2mm]
\small
  Florida State University, Tallahassee, FL 32306    \\[-2mm]
\small
  evertz@scri.fsu.edu                               \\[1mm]
\small
  $^2\,$CERN Theory Division                         \\[-2mm]
\small
  CH-1211 Gen\`eve 23, Switzerland                    \\[-2mm]
\small
  hasenbus@surya11.cern.ch                          \\[1mm]
\small
  $^3\,$School of Physics and Astronomy \\[-2mm]
\small
        Raymond and Beverly Sackler Faculty of Exact Sciences \\[-2mm]
\small
        Tel Aviv University, 69978 Tel Aviv, Israel           \\[-2mm]
\small
        marcu@vm.tau.ac.il                              \\[1mm]
\small
  $^4\,$Institut f\"ur Theoretische Physik I, Universit\"at M\"unster\\[-2mm]
\small
        Wilhelm-Klemm-Str.\ 9, W-4400 M\"unster, Germany      \\[-2mm]
\small
        pinn@yukawa.uni-muenster.de                  \\[5mm]
\normalsize  {\em Submitted to Physica A}\\[1mm]
}
\begin{document}
\maketitle %%%\vfill
\thispagestyle{empty}
\begin{abstract}\noindent \normalsize \vbox{\vspace{0mm}}\\[-4mm]
We investigate the surface width $W$ of solid-on-solid surfaces
in the vicinity of the roughening temperature
$\Tro$. Above $\Tro$, $W^2$ is expected to diverge with the system size
$L$ like $\ln L$. However, close to $\Tro$ a clean $\ln{L}$ behavior can
only be seen on extremely large lattices.
Starting from the Kosterlitz-Thouless
renormalization group, we derive an improved
formula that describes the small $L$ behavior on both sides of $\Tro$.
For the Discrete Gaussian model, we used the
valleys-to-mountains-reflections cluster algorithm in order to
simulate the fluctuating solid-on-solid surface.
The base plane above which the surface is defined is an $L \times L$
square lattice. In the simulation we took
$8\leq L\leq 256$. The improved formula fits
 the
numerical results very well. From the analysis, we estimate the
roughening temperature to be $\Tro = 0.755(3)$.
%
%%\\[2mm]\pagebreak[0]
%%\noindent {PACS} numbers:
%%        02.70+d,  % Computational Techniques
%%        05.50+q,  % Lattice theory and statistics; Ising problems
%%        68.35Rh   % Solid solid interfaces: phase trans. and crit. phenomena
\end{abstract}
\setcounter{page}{0}
\newpage
%
%***********************************************************************
\section{Introduction}

Solid-on-solid (SOS) models are useful as interface models
\cite{SOSref}. They belong to a large class of models that are believed
to be in the Kosterlitz-Thouless (KT) universality class \cite{KT}.
For SOS models, the KT transition is the roughening transition. It is
still a challenge to devise methods for the accurate study of this
transition and for unambiguous tests of the KT theory.

As a prototype of an SOS model we consider the Discrete Gaussian
(DGSOS) model, defined by the partition function
\begin{equation}\label{Zdef}
Z = \sum_{h} \, \exp \left\{ -\frac{1}{2T} \sum_{\langle i,j\rangle}
(h_i - h_j)^2 \right\} \, ,
\end{equation}
where $h_i$ are integer ``height'' variables defined on the sites $i$
of an $L\times L$ square lattice with periodic boundary conditions.  A
configuration $h$ can be viewed as a surface without overhangs, embedded
in three dimensions;
%The summation in the partition function is over all equivalence classes
%$\{h\}$ of configurations. Two configurations are considered
%equivalent if they differ by a global shift with an integer.
its energy is obtained by summing over all nearest neighbor pairs
$\langle i,j \rangle$; $T$ is the temperature (Boltzmann's constant is
set to one).  The square of the average surface width $W$ is defined
by
\begin{equation}\label{Wdef}
W^2 = \frac{1}{L^2} \sum_{i} < (h_i - h_j)^2 > \, .
\end{equation}
The model has two phases. At low temperatures the surface is smooth,
and $W$ stays finite as $L\rightarrow \infty$.  When $T$ is increased,
we encounter the roughening transition at $T=\Tro$. The KT theory
predicts \cite{SOSref,KT} that in the thermodynamic limit
\begin{equation}\label{KT1}
  W^2 \sim (\Tro-T)^{-\half} \sim \ln{\xi}
\end{equation}
as $T$ approaches $\Tro$ from below ($\xi$ is the correlation length).
For $T \geq \Tro$, $W$ diverges as $L \rightarrow \infty$. The
prediction for asymptotically large $L$ is the ``free field'' behavior
(i.e.\ Continuous Gaussian -- $h_i$ is real instead of integer)
\begin{equation}\label{KT2}
W^2 = \frac{\Teff}{\pi} \ln{L} + \mbox{const} \, .
\end{equation}
$\Teff$ is called the ``effective temperature'', and
\begin{equation}\label{KT3}
  \Teff = \frac{2}{\pi} \quad \mbox{for} \quad T=\Tro \, .
\end{equation}
Furthermore, as $T$ approaches $\Tro$ from above,
\begin{equation}\label{KT4}
 \Teff-\frac{2}{\pi} \sim (T-\Tro)^{\frac{1}{2}} \, .
\end{equation}
In principle, these formulas could be used in a numerical study in
order to verify or disprove the KT theory. In practice however this is
problematic.  In the smooth phase,
%$W^2$ grows linearly
%with exponentially increasing correlation length, and
we would need unrealistically large lattices in order to test the
power law (\ref{KT1}). This problem is related to the difficulties
encountered in the study of the dual (Villain, XY) spin models, where
it is hard to cleanly distinguish an essential singularity in the
correlation length $\xi$ (as predicted by KT) from a power law
singularity \cite{Janke,Gupta}.  In the rough phase, for large enough
temperatures, the behavior (\ref{KT2}) could be unambiguously verified
numerically \cite{ourbeta1}.  However, it turns out that close to
$\Tro$ a clean logarithm is only seen on very large lattices, and in
order to extract the values of $\Teff$ in practice we need to know the
corrections to eq.\ (\ref{KT2}). Otherwise we cannot
determine $\Tro$ by checking eq.\ (\ref{KT3}).  Furthermore,
for the largest lattice sizes accessible with
present day computers and algorithms, it turns out that eq.\
(\ref{KT4}) is not yet fulfilled for the region where eq.\ (\ref{KT2})
holds. Actually, the status of eq.\ (\ref{KT4}) is even worse, as will be
argued later.

In order to overcome these problems, we developed a renormalization
group (RG) improved formula for the dependence of $W^2$ on $L$.  This
is our main theoretical result.  The numerical part of our work shows
that the improved formula can be used for extracting $\Teff$ as close
as desired to $\Tro$, from numerical simulations on reasonably sized
lattices.  We mention that very high accuracy simulations were
possible because we have a cluster algorithm that is free of critical
slowing down (the valleys-to-mountains-reflections algorithm
\cite{VMR}).  Vectorization \cite{vectorize} also helped. From our
analysis, the best estimate for the roughening temperature is
$\Tro=0.755(3)$.

In what follows, we first derive our improved formula, then present
the analysis of the numerical results, and finally make some
additional remarks and present our conclusions.

%***********************************************************************
 \section{RG improved finite $L$ formula for the surface width}
%***********************************************************************
%

The RG flow of the DGSOS model can be described in an $x-y$ diagram
\cite{KT}. The trajectories are parametrized by $t$, the logarithm of
the changing length scale. $x(t)$ is related to the scale dependent
(``running'') temperature $T(t)$, $x(t) = \pi T(t) - 2 $, while $y(t)$
is a constant times the fugacity \cite{KT}.  The KT flow equations are
\cite{KT}:
\begin{equation}\label{KTeq}
 \begin{array}{l} \vspace{1mm} \dot y(t) = - x(t) \, y(t) \\ \dot
x(t)= - y(t)^2 \, .  \end{array}
\end{equation}
The trajectories are hyperbolas, characterized by the constant $E$
which depends on the temperature $T$ of the model (not on the running
$T(t)$ !):
\begin{equation}\label{E}
y(t)^2 - x(t)^2 = E \, .
\end{equation}
Denoting $\epsilon = \mbox{sign}(E) \sqrt{|E|}$, and $x_0=x(0)$, the
full solution of eq.\ (\ref{KTeq}) is:
\begin{equation}\label{solution}
 \begin{array}{ll} \vspace{5mm} E<0: & x(t) = \epsilon \left( 1 +
\frac{\mbox{$2\,(x_ 0-\epsilon)$}}{ \mbox{$(x_0+\epsilon)
\exp(2\epsilon\, t) - (x_0-\epsilon)$}} \right) \\ \vspace{3.5mm} E=0:
& x(t) = \;\frac{\mbox{$x_0$}}{\mbox{$1 + x_0 t$}} \\[2mm] E>0: & x(t)
= \epsilon \;\, \frac{\mbox{$x_0 - \epsilon \tan(\epsilon\,
t)$}}{\mbox{$\epsilon + x_0 \tan(\epsilon\, t) $}} \, .  \end{array}
\end{equation}
The trajectories in the rough phase reach the free field theory, and
have $\epsilon<0$; in the smooth phase they have $\epsilon>0$; at the
KT transition the critical trajectory has $\epsilon=0$ \cite{KT}.
Notice that in the rough phase $\Teff=T(t=\infty)=(2-\epsilon)/\pi$.

In order to use the RG for computing the surface width, we need to
know the contributions corresponding to each length scale. Eq.\
(\ref{Wdef}) shows that $W^2$ is a sum of a two-point-function over
all distances.  When increasing the lattice size $L$, we get
additional additive contributions from distances of order $L$.  Let us
choose
\begin{equation}\label{t}
 t = \ln{\frac{\mbox{$L$}}{\mbox{$L_0$}}} \, ,
\end{equation}
with $L_0$ some reference length scale, and let us approximate the sum
in eq.\ (\ref{Wdef}) by an integral.  For the free field theory, the
additional contributions to $W^2$ coming from an infinitesimal change
in $L$ are easily computed: since eq.\ (\ref{KT2}) is always true, with
$\Teff$ replaced by the temperature $T$, we have $\mbox{d}W^2 /
\mbox{d}t = T/\pi$.  In the case of the DGSOS model, the KT flow shows
that for trajectories in the rough phase we are close to the free
field theory if $t$ is large enough. Moreover, we are also close to the
free field theory for trajectories in the smooth phase, provided
that $L$ is much smaller than $\xi$ but still large enough.  Thus the
main contribution to $\mbox{d}W^2 / \mbox{d}t$ will be similar to the
free field case, the only (important) difference being that we replace
$T$ by the running temperature $T(t)$:
\begin{equation}\label{dWdt}
 \frac{\mbox{d}W^2}{\mbox{d}t} = \frac{T(t)}{\pi} \,.
\end{equation}
Assuming that $T(t)$ behaves according to the KT flow for length
scales larger than $L_0$, we can integrate eq.\ (\ref{dWdt}):
\begin{equation}\label{integratedWdt}
 W^2 = \frac{1}{\pi^2} \int_{0}^{t} [x(t) + 2] \, \mbox{d}t \, + \, C
\, .
\end{equation}
The constant $C$ contains the contributions of distances smaller than
$L_0$.  Using eq.\ (\ref{KTeq}) and eq.\ (\ref{E}), we can express $\mbox{d}t$
in terms of $x$ and $\mbox{d}x$,
after which eq.\ (\ref{integratedWdt}) reduces to an
elementary integral. We thus obtain our final formula for $W^2$:
\begin{equation}\label{formula}
 W^2 = \frac{2}{\pi^2} t + \frac{1}{ 2 \pi^2} \ln \left(\frac{x_0^2 +
E}{x^2 +E} \right) + \, C \, \, ,
\end{equation}
which has to be used in conjunction with eqs.\ (\ref{solution}) and
(\ref{t}).

The crucial point in the derivation of our improved formula was the
replacement, at the appropriate stage,
of the temperature $T$ with the running temperature $T(t)$.
While this is a commonly used procedure
in field theoretical RG arguments, it is not completely rigorous.
A more thorough argument, based on a block-spin calculation within
the Wilson RG framework, will be presented elsewhere \cite{elsewhere}.
%

%***********************************************************************
 \section{Simulation results}
%***********************************************************************

We performed simulations of the DGSOS model for ten different values
of the temperature $T$.  At each $T$ we considered lattice sizes of
$L=8$, 16, 32, 64, 128 and 256. Typically, we generated about
2\,000\,000 to 2\,500\,000 clusters for each temperature and lattice
size.  The expectation value of the cluster size varied in the range
$0.3 L^2$ to $0.35 L^2$. The whole project required about 400 hours on
one CRAY Y-MP processor. As will become clear from the analysis, we did
not need more than half of our runs in order to obtain the best
estimate for $\Tro$. However, our aim was also to confirm our
prediction for the $L$-dependence of $W^2$ and to determine the region
in which the improved formula is really necessary.  The simulation
results for $W^2$ are given in Table \ref{tablethick}.

As a general rule, these data are extremely well fitted by eq.\
(\ref{formula}), with fit parameters $\epsilon$, $x_0$ and $C$.  Aside
from the occasional statistical fluctuation, we did however notice
that for $T\leq 0.755$ the quality of the fits deteriorated a little.
The important results of such fits are the value of $\epsilon$, which
characterizes the trajectory, and the range of $L$ for which the fits
are good, which roughly tells us where the model enters the KT flow.
Notice that we have to decide upon a value for $L_0$. The choice of $L_0$
only affects the values $x_0$ and $C$, as can be seen after a little
algebra. Table \ref{tableepsilon} contains the fit results for
$\epsilon$, for all our values of $T$ and for various fit ranges.
Clearly, for $T\geq 0.76$ the various fit ranges give compatible
results.  In fact the data for $L=256$ hardly improve things here.
For $T\leq 0.755$ however, the fit results for $\epsilon$ sometimes
change by more than two standard
deviations if we remove the data for
$L=8$. It may be that for these temperatures the KT flow is well
reached only above $L=8$.

{F}rom Table \ref{tableepsilon} our first main numerical result strikes
the eye: since $\epsilon>0$ for $T\leq 0.75$ and $\epsilon<0$ for
$T\geq 0.76$, $\Tro$ is between $0.75$ and $0.76$. This result relies
solely on the fact that eq.\ (\ref{formula}) fits the data well, and on
eq.\ (\ref{KT3}).

%%%%%%%%%%%% TABLE 1 %%%%%%%%%%%%%%%%%%%%%%%%%%%%%%%%%%%%%%%%
\begin{table}[p]
\centering
\caption[dummy]{\label{tablethick}
%\parbox[t]{.85\textwidth}{Simulation results for $W^2$.}}
Simulation results for $W^2$.}
\vskip2ex
\small
\begin{tabular}{|c|c|c|c|c|c|c|c|}
\hline\str
$T$ & $L=8$ & $L=16$ & $L=32$ & $L=64$ & $L=128$ & $L=256$
\\[.5ex]\hline \str
0.740&0.51429(34)&0.66739(33)&0.81589(32)&0.96185(33)&1.10558(34)&1.24908(39)\\
0.745&0.51984(37)&0.67560(34)&0.82571(32)&0.97408(33)&1.12248(33)&1.26710(40)\\
0.750&0.52537(37)&0.68222(35)&0.83542(35)&0.98661(34)&1.13615(35)&1.28461(36)\\
0.755&0.53137(30)&0.69052(34)&0.84475(33)&0.99877(33)&1.15066(34)&1.30165(38)\\
0.760&0.53733(37)&0.69725(34)&0.85444(33)&1.00956(29)&1.16440(28)&1.31721(39)\\
0.770&0.54900(35)&0.71185(32)&0.87292(33)&1.03164(36)&1.18993(34)&1.34726(37)\\
0.780&0.55902(36)&0.72606(34)&0.88941(33)&1.05190(38)&1.21432(36)&1.37515(40)\\
0.800&0.58006(38)&0.75240(34)&0.92271(35)&1.09176(40)&1.26032(37)&1.42846(40)\\
0.820&0.59963(37)&0.77776(34)&0.95355(34)&1.12807(36)&1.30248(38)&1.47770(42)\\
0.850&0.62762(37)&0.81304(37)&0.99679(36)&1.18076(38)&1.36315(40)&1.54571(43)
\\[.3ex] \hline
\end{tabular}
\end{table}

In order to give a more precise determination of $\Tro$, we plotted
for each fit range $\epsilon$ versus $T$, with error bars, and
interpolated the two curves $\epsilon + \mbox{error}$ and $\epsilon -
\mbox{error}$. The intersection of the band thus obtained  with the
$\epsilon=0$ line provides an estimate of $\Tro$.  In Table
\ref{tablesummary} we collected these range-dependent estimates. They
are quite consistent with one another.  Thus, taking into account the
above observations about the quality of the fits for $T\leq 0.755$,
it would not be unreasonable to quote as our final result the value of
$\Tro$ for the $L$-range $16-256$.

%%%%%%%%%%%% TABLE 2 %%%%%%%%%%%%%%%%%%%%%%%%%%%%%%%%%%%%%%%%%%%%%%%%%%%
\begin{table}[p]
\centering
 \caption[dummy]{\label{tableepsilon} Fit
results for the parameter $\epsilon$; the first row contains the
$L$-range.}%%
\vskip2ex
\small
\begin{tabular}{|c|r|r|r|r|r|}
\hline\str
$T$ &
\mc{1}{c|}{ $8-256$} &
\mc{1}{c|}{ $16-256$} &
\mc{1}{c|}{ $32-256$} &
\mc{1}{c|}{ $8-128$} &
\mc{1}{c|}{ $16-128$}
\\[.5ex]\hline \str
0.740& 0.204(06)& 0.178(10)& 0.151(23)& 0.226(08)& 0.206(16)\\ 0.745&
0.171(07)& 0.139(14)& 0.165(21)& 0.168(11)& 0.060(57)\\ 0.750&
0.126(09)& 0.109(17)& 0.103(35)& 0.137(14)& 0.117(30)\\ 0.755&
0.030(38)& -0.078(25)& 0.059(60)& 0.054(33)& -0.102(33)\\ 0.760&
-0.124(10)& -0.130(14)& -0.128(27)& -0.131(13)& -0.151(21)\\ 0.770&
-0.211(06)& -0.210(09)& -0.221(17)& -0.211(09)& -0.205(17)\\ 0.780&
-0.277(05)& -0.287(07)& -0.278(14)& -0.278(07)& -0.300(12)\\ 0.800&
-0.387(04)& -0.387(06)& -0.388(11)& -0.386(06)& -0.387(10)\\ 0.820&
-0.478(03)& -0.484(05)& -0.494(09)& -0.471(05)& -0.474(09)\\ 0.850&
-0.601(03)& -0.599(04)& -0.590(08)& -0.603(04)& -0.601(07)
\\[.3ex] \hline
\end{tabular}
\end{table}
%

%%%%%%%%%%%% TABLE 3 %%%%%%%%%%%%%%%%%%%%%%%%%%%%%%%%%%%%%%%%%%%%%%%%%%%
\begin{table}[p]
\centering
 \caption[dummy]{\label{tablesummary} $\Tro$
from the interpolated curves $\epsilon(T)$.}
\vskip2ex
\small
\begin{tabular}{|c|c|c|c|c|c|}
\hline\str
 fit range & $8-256$ & $16-256$ & $32-256$ & $8-128$ & $16-128$
\\[.5ex]\hline \str
estimated
$\Tro$&0.7555(25)&0.7535(15)&0.7550(30)&0.7555(25)&0.7515(55)
\\[.3ex] \hline
\end{tabular}
\end{table}

For a more conservative estimate of $\Tro$ we plotted the values of
$\epsilon$ from the ranges $8-256$ and $16-256$, together with their
error bars, on the same plot. We interpolated the upper and lower
envelopes of the error bars. From the intersection of the band thus obtained
with the $\epsilon=0$ line we get the estimate $\Tro = 0.755(3)$.
Notice that the best estimate in the literature \cite{KTmatch},
$\Tro=0.7524(7)$, was obtained by a completely different method
(matching with the critical block spin flow of the BCSOS model), that
does not test directly any of the formulas derived from the KT theory.
The best estimate by other authors \cite{Janke}, $\Tro=0.752(5)$ (from
the analysis of the correlation length and susceptibility in the
massive phase of the Villain model), is also consistent with the
result presented here.

At the beginning of section 2 we explicitly wrote down the $t$
dependence of the running temperature $T(t)$.
With the numerically determined fit parameters
$\epsilon$ and $x_0$, we can thus compute the flow of $T(t)$
numerically. If we use $x(t)$ instead of $T(t)$, we can neatly
plot the points $(x(t),y(t))$ inside the standard KT flow diagram.
We can now do the following consistency check.
The differences $\pi \, \Delta W^2 / \Delta t = (\pi/\ln2)\,[W^2(2L)-W^2(L)]$,
shown in Table \ref{tablelog}, should
be discrete approximants of $T(t)$,
by eq.\ (\ref{t}) and (\ref{dWdt}).
Thus if we again plot the values of the points $(x(t),y(t))$,
this time using the discrete approximation,
we expect to obtain a similar diagram.
We did this exercise, and indeed the two diagrams were
almost identical.

In the last column of Table \ref{tablelog} we show the
values of $\Teff=(2-\epsilon)/\pi$, obtained by again taking for each
$T>\Tro$ the envelope of the error bars from the fits with $L$-ranges
$8-256$ and $16-256$. Above $\Tro$, if $L$ is large enough for
eq.\ (\ref{KT2}) to hold, the running temperature stabilizes to the value
$\Teff$. By looking at how the the results in the rows of
Table \ref{tablelog} stabilize, we see that our data for $W^2$
enter the asymptotic regime of eq.\ (\ref{KT2})
for $0.8\leq T\leq 0.85$ clearly, and
for $T=0.78$ just barely. For $\Tro\leq 0.77$ however,
this is far from being true, even at
$L=256$. Notice that in order to understand the validity region of
eq.\ (\ref{KT2}) we did
need the values of $W^2$ for $L=256$. More importantly,
notice that our results show that the use of eqs.\ (\ref{KT2}) and (\ref{KT3})
for determining $\Tro$ (like e.g.\ in \cite{Weeks,Kleinert}) leads
to a consistent underestimate.

%%%%%%%%%%%% TABLE 4 %%%%%%%%%%%%%%%%%%%%%%%%%%%%%%%%%%%%%%%%%%%%%%%%%%%
\begin{table}
 \caption[dummy]{\label{tablelog}
  The differences $\pi \, \Delta W^2 / \Delta t$ compared to $\Teff$}%
 \centering
\vskip2ex
\small
\begin{tabular}{|c|c|c|c|c|c|c|}
\hline\str
$T$  & $8-16$   & $16-32$   & $32-64$ & $64-128$ & $128-256$ & $\Teff$
\\[.5ex]\hline \str
0.740& 0.6939(21)& 0.6731(21)& 0.6615(21)& 0.6514(22)& 0.6504(24)& $T<\Tro$ \\
0.745& 0.7059(23)& 0.6804(21)& 0.6724(21)& 0.6726(21)& 0.6555(23)& $T<\Tro$ \\
0.750& 0.7109(23)& 0.6944(22)& 0.6852(22)& 0.6778(22)& 0.6729(23)& $T<\Tro$ \\
0.755& 0.7213(21)& 0.6990(22)& 0.6981(21)& 0.6884(21)& 0.6843(23)&
$T\approx\Tro$\\
0.760& 0.7248(23)& 0.7124(21)& 0.7031(20)& 0.7018(18)& 0.6926(22)& 0.6777(48)\\
0.770& 0.7381(21)& 0.7300(21)& 0.7194(22)& 0.7174(22)& 0.7131(23)& 0.7035(29)\\
0.780& 0.7571(22)& 0.7404(22)& 0.7365(23)& 0.7361(24)& 0.7290(24)& 0.7267(35)\\
0.800& 0.7811(23)& 0.7719(22)& 0.7662(24)& 0.7640(25)& 0.7620(25)& 0.7598(19)\\
0.820& 0.8074(23)& 0.7967(22)& 0.7910(22)& 0.7905(24)& 0.7942(26)& 0.7900(22)\\
0.850& 0.8404(24)& 0.8328(23)& 0.8338(24)& 0.8267(25)& 0.8274(27)& 0.8273(16)
\\[.3ex] \hline
\end{tabular}
\end{table}

In order to test eq.\ (\ref{KT4}), we fitted the values for $\Teff$ from
Table \ref{tablelog}. The fit was not at all good.
We then allowed for a free power instead of the power $\frac{1}{2}$.
The fit was now good, but the power was 0.60(4). Disregarding the
point farthest away from $\Tro$, $T=0.85$,
the situation did not improve: the power changed to 0.62(7).
The fitted value for $\Tro$ was 0.753(2) with the point $T=0.85$ included,
and 0.752(4) without it. While these values for $\Tro$ are reasonable, the
fact remains that the power $\frac{1}{2}$ is not yet seen even as close to
$\Tro$ as our data in the rough phase are. Notice that this conclusion
implies in particular that we cannot use
eq.\ (\ref{KT4}) in order to fit results in a region where the
simple behavior (\ref{KT2}) applies on lattices
of still manageable size.

As a last issue, let us remark that in the absence of a theory, one may
be simply tempted to make some ``reasonable'' ansatz  for the corrections
to eq.\ (\ref{KT2}). We tried to fit the data with a $\ln\ln L$ correction
(the coefficient in front of $\ln\ln L$ is the third fit parameter
besides $\Teff$ and the constant).
The fits were as good as those with eq.\ (\ref{formula}), if not better.
However, the values of $\Teff$ thus obtained were clearly wrong.
It is not difficult to understand the numerics behind this phenomenon.
The main point is, however, to view this as another example of the
danger of analyzing simulation results without a solid theoretical
basis.

Along the same lines, let us remark that we found a different modification
of eq.\ (\ref{KT2}) to also fit the data very well: instead of taking
$\ln L$ we took a power of $\ln L$ (this power is the third fit parameter).
The power that allowed for good fits very close to $\Tro$ never deviated
from the value 1 by more than 10\%. Nevertheless, the fitted values for
$\Teff$ were again clearly wrong with this procedure.

%%%%%%%%%%%%%%%%%%%%%%%%%%%%%%%%%%%%%%%%%%%%%%%%%%%%%%%%
\section{Conclusions}
%%%%%%%%%%%%%%%%%%%%%%%%%%%%%%%%%%%%%%%%%%%%%%%%%%%%%%%%

We have derived a renormalization group improved formula for the finite
size behavior of the SOS surface width in the vicinity of the roughening
transition.
The improved formula was tested in a high accuracy simulation of the
DGSOS model, and found to describe the data excellently.
As a result of our analysis, we:
\begin{itemize}
\begin{itemize}
\item verified an important aspect of the KT scenario;
\item gave a precise determination of $\Tro$;
\item found the region in which eq.\ (\ref{KT2}) cannot be used
      unless we consider much larger lattice sizes;
\item found that the region of applicability of eq.\ (\ref{KT4}) is much
      smaller than previously assumed.
\end{itemize}
\end{itemize}

\section*{Acknowledgements}
This work was supported in part by the Deutsche Forschungsgemeinschaft
(DFG), the German-Israeli Foundation for
Research and Development (GIF) and by the Basic Research Foundation of
the Israel Academy of Sciences and Humanities,
as well as the Florida State University Supercomputer Computations Research
Institute which is partially funded by the U.S. Department of Energy.
We would like to express
our gratitude to the HLRZ at KFA J\"ulich where most of our computer
runs were performed.

%%%%%%%%%%%%%%%%%%%%%%%%%%%%%%%%%%%%%%%%%%%%%%%%%%%%%%%%%%%%%%%%%%%%%%%%%%%%
%     BIBLIOGRAPHY
%%%%%%%%%%%%%%%%%%%%%%%%%%%%%%%%%%%%%%%%%%%%%%%%%%%%%%%%%%%%%%%%%%%%%%%%%%%%
%   \pagebreak% [3]

\end{document}